\documentclass[12pt]{article}
\usepackage{graphicx}
\usepackage{hyperref}
\usepackage{subfigure}
\usepackage{amsmath}

\newcommand\pubdate{\today}

\def\columbia{Department of Physics\\ Columbia University, 538W 120th Street, New York, NY, 10027, USA}
\def\torvergata{Dipartimento  di  Fisica  and  INFN \\ Universit\`{a} di Roma `Tor Vergata', Via della Ricerca Scientica 1, I-00133 Roma, Italy}
\def\sapienza{Dipartimento  di  Fisica  and  INFN \\  `Sapienza' Universit\`{a} di Roma, P.le Aldo Moro 5, I-00185 Roma, Italy}
\usepackage{color}

\def\Title#1{\begin{center} {\Large #1 } \end{center}}
\def\Author#1{\begin{center}{ \sc #1} \end{center}}
\def\Address#1{\begin{center}{ \it #1} \end{center}}

\newcommand\pubblock{\rightline{\begin{tabular}{l} \\
         \pubdate  \end{tabular}}}
\newenvironment{Abstract}{\begin{quotation}  }{\end{quotation}}
\newenvironment{Presented}{\begin{quotation} \begin{center} 
             PRESENTED AT\end{center}\bigskip 
      \begin{center}\begin{large}}{\end{large}\end{center} \end{quotation}}
\def\Acknowledgements{\bigskip  \bigskip \begin{center} \begin{large}
             \bf ACKNOWLEDGEMENTS \end{large}\end{center}}

\def\beq{\begin{equation}}
\def\eeq#1{\label{#1}\end{equation}}
\def\eeqn{\end{equation}}

\def\beqa{\begin{eqnarray}}
\def\eeqa#1{\label{#1}\end{eqnarray}}
\def\eeqan{\end{eqnarray}}

\let\littlebar=\bar
\let\bar=\overbar

\def\Dslash{\not{\hbox{\kern-4pt $D$}}}
\def\dslash{\not{\hbox{\kern-2pt $\del$}}}

\def\msb{{\bar{\ssstyle M \kern -1pt S}}}

\begin{document}
\begin{titlepage}
\pubblock

\vfill
\Title{The $Z_c^{(\prime)}\to\eta_c\rho$ decay as a discriminant between tetraquarks and meson molecules}
\vfill
\Author{Angelo Esposito\footnote{Speaker.}}
\Address{\columbia}
\Author{Andrea L. Guerrieri}
\Address{\torvergata}
\Author{Alessandro Pilloni}
\Address{\sapienza}
\vfill
\begin{Abstract}
Understanding the nature of the exotic $XYZ$ resonances is one of the open problems in hadronic spectroscopy. Despite the experimental efforts, the structure of these particles still lacks of an accepted theoretical framework. We propose to use the $Z_c^{(\prime)}\to\eta_c\rho$ decays as a possible discriminant between two of the most popular models: the compact tetraquark and the loosely bound meson molecule. We show that the predictions obtained within the two pictures are significantly different and therefore the proposed decay channel might shed some light on the nature of these states.
\end{Abstract}
\vfill
\begin{Presented}
The 7th International Workshop on Charm Physics (CHARM 2015)\\
Detroit, MI, 18-22 May, 2015
\end{Presented}
\vfill
\end{titlepage}
\def\thefootnote{\fnsymbol{footnote}}
\setcounter{footnote}{0}
%

\section{Introduction}

\begin{figure}
\centering
\subfigure[]{
\includegraphics[width=0.47\textwidth]{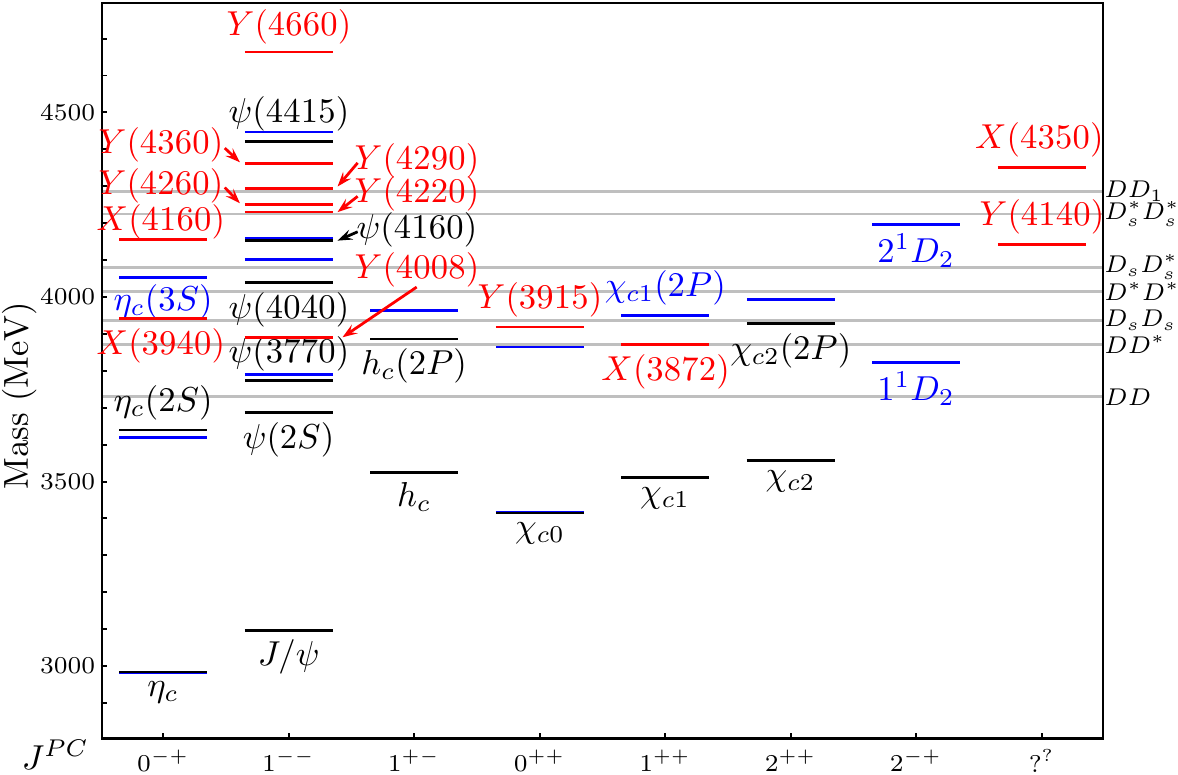}
}
\subfigure[]{
\includegraphics[width=0.47\textwidth]{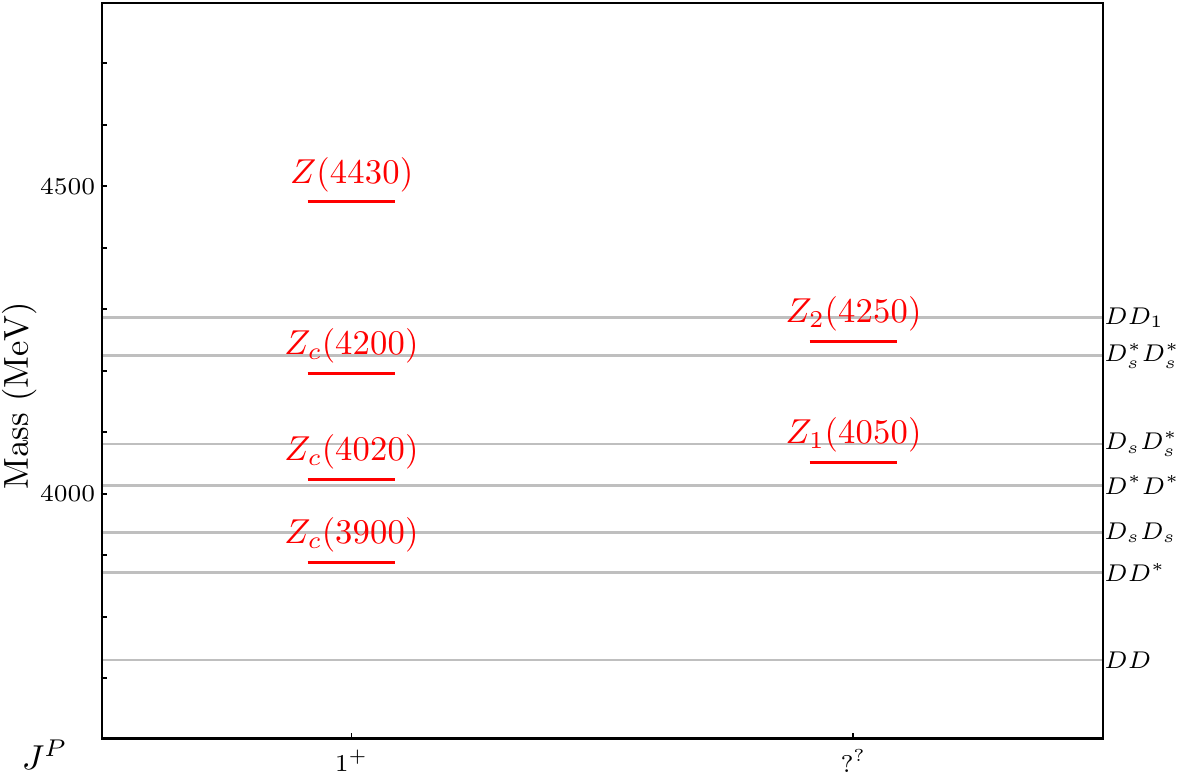}
}
\caption{Spectrum of the charmonium and charmonium-like states. Left panel: black and blue lines are the observed and expected charmonia respectively~\cite{review}. Red lines are the observed neutral exotic states. Right panel: observed charged exotic states. On the right of each panel we also report the open charm thresholds.} \label{states}
\end{figure}

In the past decade, several experiments observed a set of charmonium-like resonances sharing similar ``exotic'' properties --- see~\cite{review} for a review. The spectrum of these particles, the so-called $XYZ$ states, is reported in Fig.~\ref{states}. In particular, the observation of charged resonances decaying into charmonia is a compelling evidence for the existence of four-quark states. The most popular models proposed to describe the internal structure of these particles are the compact diquark-antidiquark (or just tetraquark from now on)~\cite{maianiX,maianitypeII}, the loosely bound meson molecule~\cite{mols,cleven}, the so-called hadro-charmonium~\cite{hadrocharmonium} and the gluonic hybrid~\cite{hybrids}. None of these models has been generally accepted as the right one yet. 

Despite the lack of observation of many of the states predicted by the tetraquark model, the recent discovery of two baryon-like resonances with opposite parities decaying into $J/\psi\,p$~\cite{lhcb} has given new interest to the diquark models~\cite{penta}.

In this work we summarize what developed in~\cite{etac}, focusing our attention on two charged states: the $Z_c(3900)$ and the $Z_c^\prime(4020)$. The first one has been observed by BES and Belle in the $J/\psi\,\pi^+$~\cite{zc} final state and by BES in the ${(D\littlebar{D}^*)}^+$~\cite{zcDDstar} channel\footnote{Here and in what follows charged conjugate modes are undestood, unless otherwise specified.}, while the latter one has been observed by BES in $h_c\,\pi^+$ and ${(D^*\littlebar{D}^*)}^+$~\cite{zc4020}. For both of them the most likely quantum numbers are $(I^G)J^{PC}=(1^+)1^{+-}$, where the charge conjugation refers to the eigenvalue of the neutral isospin partner.

Similarly to what happens to many other exotic resonances, the $Z_c$ and $Z_c^\prime$ lie very close in mass to the $D\littlebar D^*$ and $D^*\littlebar D^*$ thresholds respectively (see again Fig.~\ref{states}) and therefore have been interpreted as loosely bound molecules~\cite{riddle}, despite being slightly above threshold. A $J^{PC}=1^{+-}$ state with mass around $3882$ MeV is however also predicted by the constituent diquark-antidiquark model~\cite{maianiX}, together with its radial excitation around $4470$ MeV. These states have been identified respectively with the $Z_c(3900)$ and with the recently discovered $Z(4430)$~\cite{z4430}. While in the first version of the tetraquark model (the so-called ``type~I'') the $Z_c^\prime(4020)$ was not included, it is now nicely accomodated by a recent ``type~II'' paradigm~\cite{maianitypeII}.

Here we show how the $Z_c^{(\prime)}\to\eta_c\rho$ decay channel can be used as a tool to differentiate between two of the possible internal structures of these charged resonances. For the $Z_c(3900)\to \eta_c\rho$ process, some considerations about the consequences of the heavy quark spin symmetry within the molecular and tetraquark pictures have already been explored in~\cite{Voloshin}.

\section{Compact tetraquark}
In the constituent diquark-antidiquark model the Hamiltonian that describes the interaction between the components of the state is given by:
\begin{align} \label{eq:H}
H=\sum_i m_i-2\sum_{a,i\neq j}\kappa_{ij}\vec{S}_i\cdot\vec{S}_j\frac{\lambda^a_i}{2}\frac{\lambda^a_j}{2},
\end{align}
where $m_i$ are the masses of the constituents, $\kappa_{ij}$ are unknown couplings, $\vec{S}_i$ are spin vectors and $\lambda_i^a$ are the Gell-Mann matrices. In the type~I model~\cite{maianiX}, the couplings are extracted from the spectrum of ordinary mesons and baryons and the mass of the $X(3872)$ is used as an input to determine the diquark mass, $m_{[cq]}$. In the type~II~\cite{maianitypeII}, instead, since the spin-spin interaction is a contact term, it is assumed that the only relevant interactions are those within the diquarks and hence all the couplings are set to zero except for $\kappa_{cq}=\kappa_{\littlebar c\littlebar q}$. Depending on which of the two ansatz is chosen, the physical states $X(3872)$, $Z_c(3900)$ and $Z_c^\prime(4020)$ are identified with different combinations of the eigenstates of the Hamiltonian in~\eqref{eq:H}.

As far as the decays of tetraquarks are concerned one can instead resort to the well-known heavy quark spin symmetry~\cite{hqss} to write the amplitudes for the decay into charmonia as a Clebsch-Gordan spin factor times a transition matrix element~\cite{etac}. This is valid up to corrections of order $\Lambda_{QCD}/m_c\simeq25\%$, where $m_c\simeq 1.5\text{~GeV}$ is the constituent charm quark mass.

For the processes of interest, the most general Lorentz-invariant matrix elements that behave properly under parity and charge conjugation are
\begin{subequations}
\begin{gather}
 \left\langle J/\psi\left(\eta,p\right)\,\pi\left(q\right) | Z\left(\lambda,P\right) \right\rangle = g_{Z\psi\pi}\, \lambda \cdot \eta,\quad \left\langle \eta_c\left(p\right)\,\rho\left(\epsilon,q\right) | Z\left(\lambda,P\right) \right\rangle = g_{Z\eta_c\rho} \,\lambda \cdot \epsilon,\\  \left\langle h_c\left(p,\eta\right)\,\pi\left(q\right) | Z\left(\lambda,P\right) \right\rangle = \frac{g_{Z h_c\pi}}{M_Z^2}\, \epsilon^{\mu\nu\rho\sigma}\lambda_\mu \eta_\nu P_\rho q_\sigma,
\end{gather}
\end{subequations}
where $\lambda$, $\eta$ and $\epsilon$ are polarization vectors, $p$, $q$ and $P$ are four-momenta and the $g$s are effective couplings with dimension of a mass.

Since we have no information on the couplings, some kind of ansatz is required here as well. To test the degree of model dependence of our calculation we make two different assumptions: \emph{(a)} we neglect the spatial dependence of the wave functions and hence assume that the couplings are universal, the differences between the different matrix elements being only of kinematical nature; \emph{(b)} we use a dynamical model recently developed in~\cite{dynamics}. In this picture the couplings for the decays into charmonia are proportional to the propabability density of the charmonium itself computed at a distance $r_Z$. The last one is defined as the classical turning point of a diquark-antidiquark pair moving away from each other and interacting via a spinless Cornell potential.

In Tab.~\ref{tab:tetra} we report the predictions obtained within the tetraquark model.
 \begin{table}[h]
\centering
 \begin{tabular}{l|c|c|c|c}\hline\hline
 & \multicolumn{2}{c|}{Kinematics only} & \multicolumn{2}{c}{Dynamics included} \\ \hline
  & type~I & type~II & type~I & type~II \\ \hline\hline
\rule{0pt}{20pt} $\displaystyle{\frac{\mathcal{BR}\left(Z_c \to \eta_c\,\rho\right)}{\mathcal{BR}\left(Z_c \to J/\psi\,\pi\right)}}$ & $\left(3.3^{+7.9}_{-1.4}\right)\times 10^2$ & $0.41^{+0.96}_{-0.17}$ & $\left(2.3^{+3.3}_{-1.4}\right)\times10^2$ & $0.27^{+0.40}_{-0.17}$ \\[10pt] \hline
\rule{0pt}{20pt} $\displaystyle{\frac{\mathcal{BR}\left(Z_c^\prime \to \eta_c\,\rho\right)}{\mathcal{BR}\left(Z_c^\prime \to h_c \pi\right)}}$ & \multicolumn{2}{c|}{$\left(1.2^{+2.8}_{-0.5}\right)\times10^2$} & \multicolumn{2}{c}{$6.6^{+56.8}_{-5.8}$} \\[10pt] \hline 
 \end{tabular}
 \caption{Predicted ratios of branching fractions for the $Z_c^{(\prime)}$ states according to the main tetraquark models. The first and second columns are computed under the assumptions \emph{(a)} and \emph{(b)} respectively, as explained in the text. Both type~I and type~II models give the same predictions for the $\mathcal{BR}\left(Z_c^\prime \to \eta_c\,\rho\right)/\mathcal{BR}\left(Z_c^\prime \to h_c \pi\right)$, since both $h_c$ and $\eta_c$ have spin $s_{c\littlebar c}=0$. The errors are estimated via a toy MC simulation.}
\label{tab:tetra}
 \end{table}

\section{Loosely bound molecule}
\begin{figure}[t]
\centering
\subfigure[]{ \label{fig:1a}
\includegraphics{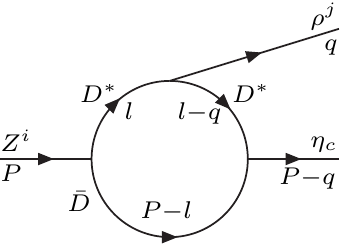}
}
\hspace{.5em}
\subfigure[]{ \label{fig:1b}
\includegraphics{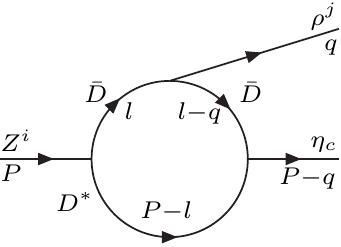}
}
\hspace{.5em}
\subfigure[]{ 
\label{fig:1c}
\includegraphics{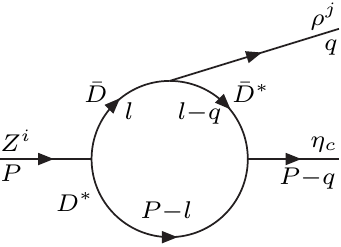}
}
\subfigure[]{ \label{fig:1a_prime}
\includegraphics{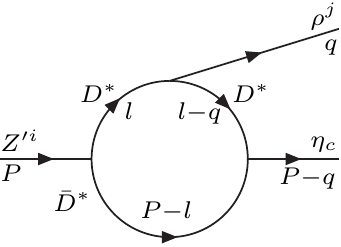}
}
\hspace{.5em}
\subfigure[]{ \label{fig:1b_prime}
\includegraphics{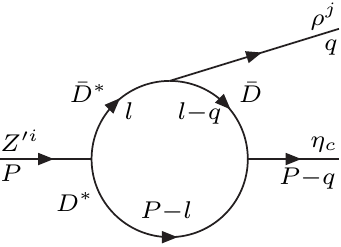}
}
\caption{Possible one-loop diagrams for the $Z_c$ (upper figures) and the $Z_c^\prime$ (lower figures) decaying into $\eta_c\,\rho$. The charge conjugate diagrams are omitted.} \label{fig:diagr}
\end{figure}

As already mentioned, in the molecular picture the $Z_c^{(\prime)}$ is interpreted as a $D^{(*)}\littlebar D^*$ loosely bound state. The interaction between the exotic particles and the heavy and light mesons is commonly described by means of the so-called Non Relativistic Effective Field Theory (NREFT)~\cite{cleven}. This is a non-relativistic limit of the Heavy Quark Effective Theory (HQET) together with the Chiral Effective Field Theory ($\chi$EFT). The complete Lagrangian of interest for our study is fully reported in~\cite{etac}, together with the choice of couplings for the interaction between the different fields. The term describing the interaction between the $Z_c^{(\prime)}$ and the charmed mesons is given by
\begin{equation}
\mathcal{L}_{Z_c^{(\prime)}}=\frac{z^{(\prime)}}{2}\left\langle \mathcal{Z}^{(\prime)}_{\mu,ab} \littlebar H_{2b}\gamma^\mu\littlebar H_{1a}\right\rangle+h.c.,
\end{equation}
where $\mathcal{Z}^{(\prime)}_{\mu,ab}$ and $\littlebar H_{ia}$ are the HQET fields for the doubly heavy $Z_c$'s states and for the $D$ mesons respectively. The $z^{(\prime)}$ are, instead, unknown effective couplings. See again~\cite{etac} for details and definitions.
In principle, such an effective theory is a valid description of the decays of the $Z_c^{(\prime)}$ regardless of its internal structure since the form of the interaction is only dictated by symmetry considerations. The molecular nature of a state is imposed by forcing it to couple to its own constituents only. Therefore, the decays into final states different from the latter ones (charmonia in our case) can only happen via heavy meson loops. The most relevant one-loop diagrams for the $Z_c^{(\prime)}\to\eta_c\rho$ process are reported in Fig.~\ref{fig:diagr}.

Moreover, since the molecular states are near threshold, the velocities of the mesons inside the loops are given by $v_X\simeq\sqrt{|M_Z-2M_D|/M_D}$ and are typically small. This allows to perform a power counting procedure in order to estimate the relevance of a certain diagram~\cite{cleven}. Using this technique, one finds that the omission of diagrams with more than one loop introduces a $15\%$ relative error on each single amplitude.

Given the previous set up, the predictions obtained within the meson molecule framework are:
\begin{align}
\frac{\mathcal{BR}(Z_c\to\eta_c\,\rho)}{\mathcal{BR}(Z_c\to J/\psi\,\pi)} = \left(4.6_{-1.7}^{+2.5}\right)\times 10^{-2}\,;\quad\frac{\mathcal{BR}(Z_c^\prime\to\eta_c\,\rho)}{\mathcal{BR}(Z_c^\prime\to h_c\,\pi)}=\left(1.0_{-0.4}^{+0.6}\right)\times 10^{-2}\,.
\end{align}

As an additional result one can also assume that the total width of the $Z_c^{(\prime)}$ is saturated by the $D^{(*)}\littlebar D^*$, $\eta_c\,\rho$, $h_c\,\pi$, $J/\psi\,\pi$ and $\psi(2S)\,\pi$ final states and therefore fit the couplings to the constituents from the experimental data. This gives
\begin{align}
\left|\,z\,\right|=\left(1.26_{-0.14}^{+0.14}\right) \text{ GeV}^{-1/2}\quad\text{ and }\quad\left|\,z^\prime\,\right|= \left(0.58_{-0.19}^{+0.22}\right) \text{ GeV}^{-1/2}.
\end{align}
Once these couplings are given one can also make the following predictions for the comparison between the two charged resonances decaying into the same final states:
\begin{align} \label{eq:extra}
\frac{\mathcal{BR}(Z_c\to h_c\,\pi)}{\mathcal{BR}(Z_c^\prime\to h_c\,\pi)}=0.34_{-0.13}^{+0.21}\,;\quad\frac{\mathcal{BR}(Z_c\to J/\psi\,\pi)}{\mathcal{BR}(Z_c^\prime\to J/\psi\,\pi)}=0.35_{-0.21}^{+0.49}\,.
\end{align}
 
\section{Conclusions}
\begin{figure}[t]
\centering
\includegraphics[width=.49\textwidth]{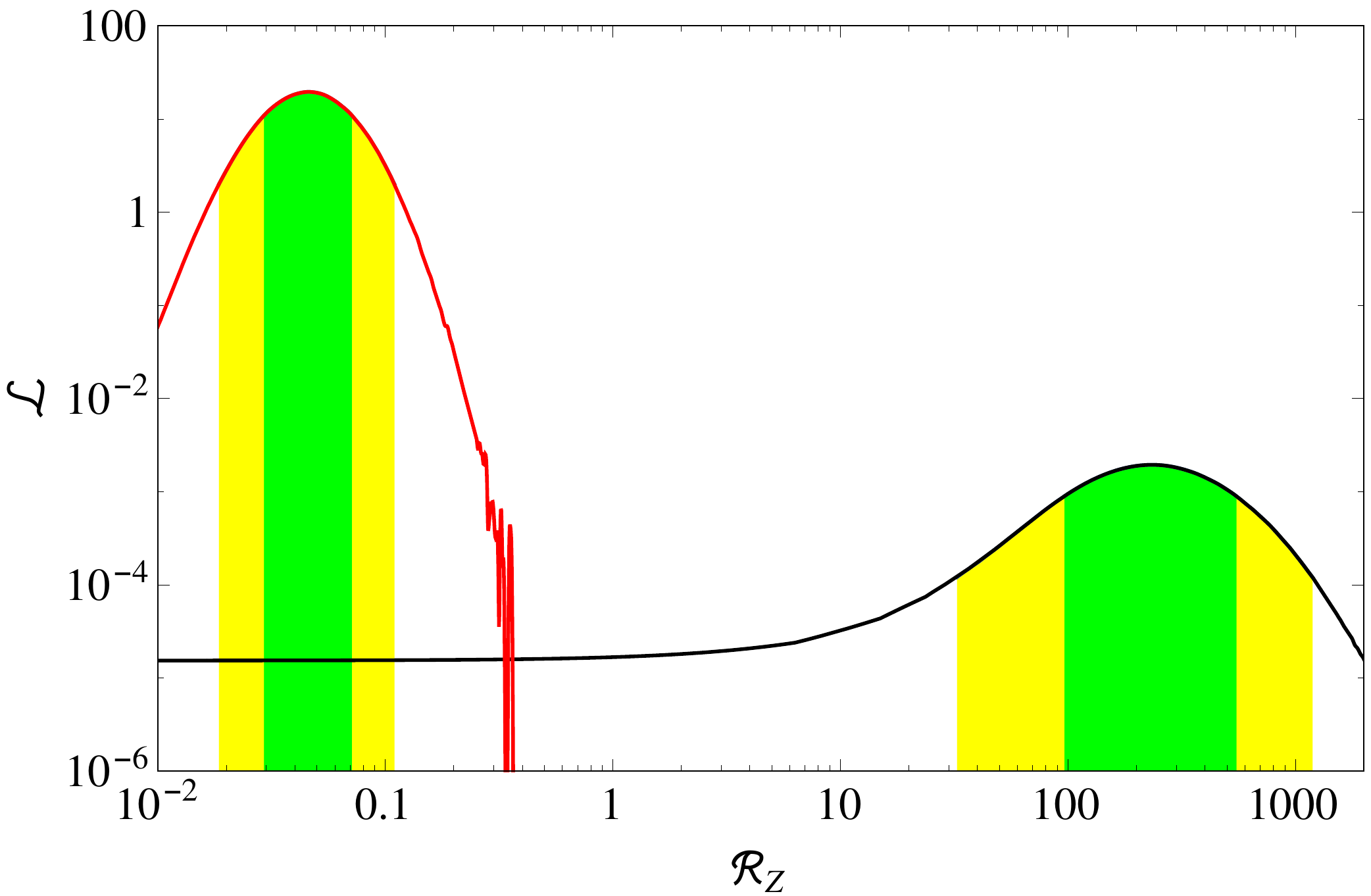}
\includegraphics[width=.49\textwidth]{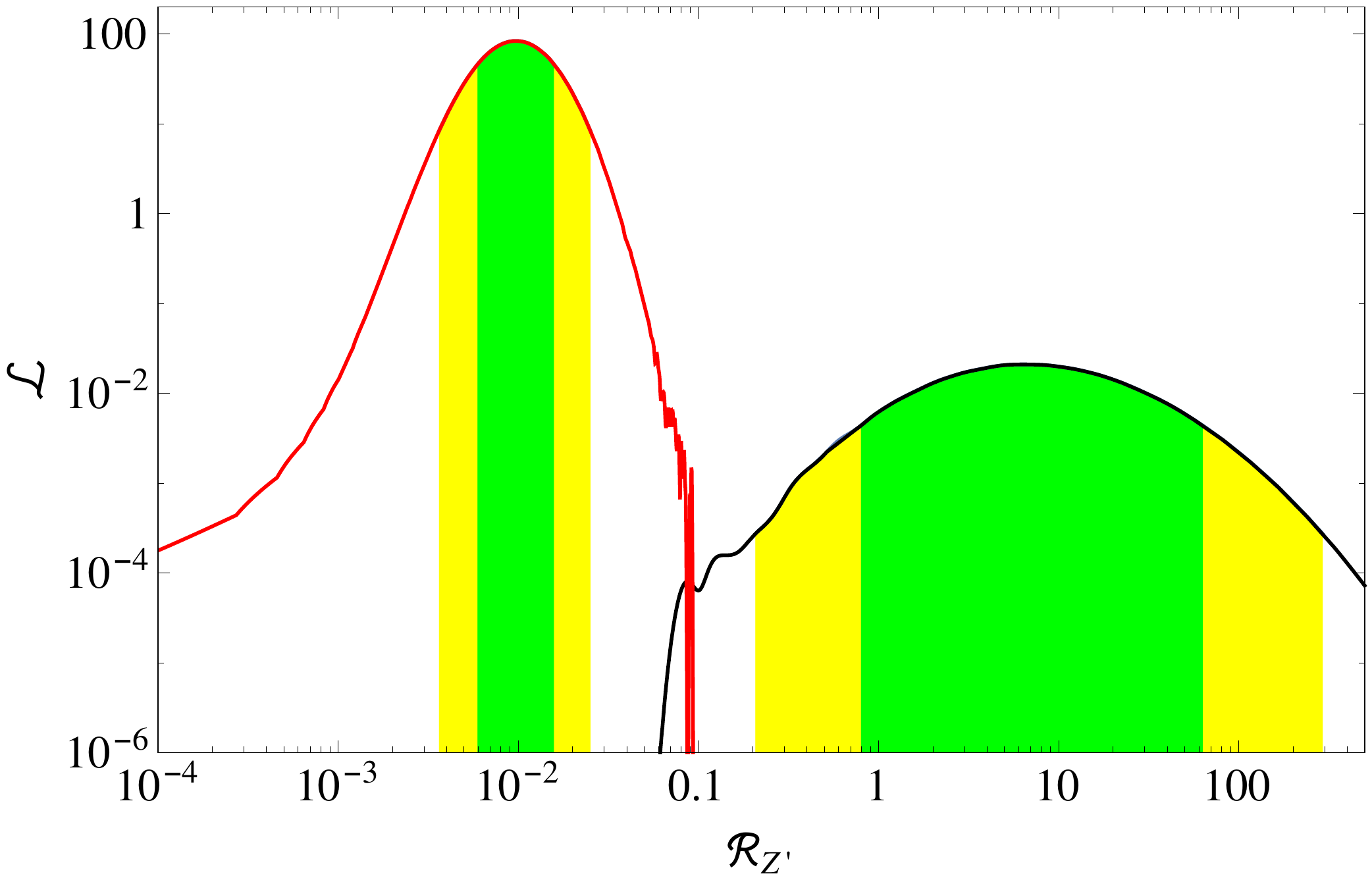}
\caption{Likelihood curves for $\mathcal{BR}(Z_c\to \eta_c\,\rho)/\mathcal{BR}(Z_c\to J/\psi\,\pi)$ (left) and $\mathcal{BR}(Z_c^\prime\to \eta_c\,\rho)/\mathcal{BR}(Z_c^\prime\to h_c\,\pi)$ (right). The red curve is the molecular prediction, whereas the black one gives the predictions for the dynamical type~I tetraquark model. The green (yellow) bands represent the $68\%$ ($95\%$) confidence region.}
\label{fig:likelihood}
\end{figure}

We can now properly compare the predictions obtained within the two models presented --- see Fig.~\ref{fig:likelihood}. As one can see, according to the dynamical type~I tetraquark model the $Z_c\to\eta_c\,\rho$ decay should be enhanced with respect to the already observed $Z_c\to J/\psi\,\pi$. The opposite is expected in the meson molecule picture and the two predictions are separated by more than $2\sigma\,(95\%\text{ C.L.})$. A similar thing holds for the $Z_c^\prime\to\eta_c\,\rho$ decay with respect to the $Z_c^\prime\to h_c\,\pi$ one. In the last case, however, the predictions for the type~I and type~II models are the same and hence the result is more model independent. The values obtained under the assumption of no dynamics for the tetraquark turn out to be even more separated from the molecular ones. For the $Z_c$ in the type~II paradigm, instead, the two models give predictions which are compatible within $2\sigma$.

Lastly, the results reported in Eq.~\eqref{eq:extra} show that in the molecular picture one expects $\mathcal{BR}(Z_c\to h_c\,\pi)/\mathcal{BR}(Z_c^\prime\to h_c\,\pi)<0.88$ and $\mathcal{BR}(Z_c\to J/\psi\,\pi)/\mathcal{BR}(Z_c^\prime\to J/\psi\,\pi)<1.86$ at $95\%$ C.L.. This means that the two charged resonances should be seen in both the $h_c\,\pi$ and $J/\psi\,\pi$ final states with comparable rates. While this seems to agree with the data in the first case, where a small hint of $Z_c$ is seen, it might be at odds with the experiments for the $J/\psi\,\pi$ channel, where no $Z_c^\prime$ has been observed so far.

In conclusion, we showed how the analysis of the $Z_c^{(\prime)}\to\eta_c\,\rho$ decay can be used as a probe of the internal structure of these charged states and hence provide a tool to discriminate between two of the most accepted models for the exotic $XYZ$ mesons. Experimental data on this channel could therefore shed some light on the now long-standing question about the nature of the $Z_c$ and $Z_c^\prime$ resonances.

\Acknowledgements
We are grateful to F.~Piccinini and A.~D.~Polosa for many useful discussions and the fruitful collaboration.


\end{document}